\documentclass[aps,twocolumn,prl,showpacs,color,psfig,epsf]{revtex4}

\usepackage{amsmath}

\usepackage{amsfonts}

\usepackage{graphicx}
\usepackage{ulem}

\begin{document}

\title{Facilitated diffusion on mobile DNA: configurational traps and sequence heterogeneity}

\author{C. A. Brackley, M. E. Cates, D. Marenduzzo}
\affiliation{SUPA, School of Physics and Astronomy, University of 
Edinburgh, Mayfield Road, Edinburgh, EH9 3JZ, UK}

\begin{abstract}
We present Brownian dynamics simulations of the facilitated diffusion
of a protein, modelled as a sphere with a binding site on its surface, 
along DNA, modelled as a semi-flexible
polymer. We consider both the effect of DNA organisation
in 3D, and of sequence heterogeneity. We find that in a network of DNA loops, as
are thought to be present in bacterial DNA, the search process is very
sensitive to the spatial location of the target within such loops. Therefore,
specific genes might be repressed or promoted by changing the local
topology of the genome. On the other hand, sequence heterogeneity creates traps 
which normally slow down facilitated diffusion. When suitably 
positioned, though, these traps can, surprisingly, render the search process 
much more efficient.
\end{abstract}

\maketitle


In living cells, proteins routinely need to reach a target 
positioned on the DNA, e.g. to initiate transcription of one gene, 
or to silence or suppress another. 
Importantly, the search for the target has to be both rapid and
efficient. Most experimental results suggest that, within 
bacterial cells, this process takes about two orders of magnitude less 
time than one would 
estimate by assuming unbiased 3D protein diffusion~\cite{Riggs,Xie,Xie2}.
How is such an efficient search realised in practice? 
The commonly accepted theory is that when seeking their target, 
proteins alternate between phases of free diffusion through the cytoplasm, 
and phases in which they slide along the DNA, 
effectively performing 1D diffusion along its backbone~\cite{Wang,Gowers,Elf}.
This combined strategy is known as ``facilitated diffusion''~\cite{Berg,Marko,Loverdo,Voituriez,Voituriez2,Metzler}.

A simple scaling argument~\cite{Marko} to predict the 
magnitude of the mean search time, $\tau_s$, that a protein needs to
find a target on the DNA, is as follows. The key parameters 
are the DNA length $L$, the volume of the cell $V$, the 3D and 1D
diffusion coefficients, respectively $D_3$ and $D_1$ (experiments suggest
$D_1<D_3$~\cite{D_3}), and, crucially, the ``sliding length'', $l_s$.
This is defined as the typical
length of DNA which the protein explores during one episode of 1D diffusion. 
Via dimensional analysis, one can estimate a typical
time spent on a 3D excursion as
$\tau_{\mathrm{3D}}\sim V/D_3 L$, while a typical sliding time is 
$\tau_{\mathrm{1D}}\sim l_s^2/D_1$. Furthermore, the mean number of 1D-3D search 
rounds is $N_{s}\sim L/l_s$~\cite{notesearchrounds}. 
One can combine these formulae to estimate $\tau_s$ 
by summing the time spent performing 3D and 1D diffusion, 
\begin{equation}\label{theory}
\tau_s = N_s (\tau_{\mathrm{1D}}+\tau_{\mathrm{3D}}) \sim 
A \frac{V}{D_3 l_s} + B \frac{L l_s}{D_1},
\end{equation}
where $A$ and $B$ are geometry-dependent constants which cannot be inferred 
from simple scaling~\cite{Langowski,Marko}. The most important result from
the theory is that there is an optimal sliding length which minimises
$\tau_s$, given by $l_s^*=\sqrt{(AD_{1}V)/(BD_{3}L)}$. With
typical parameters for bacteria and assuming $A\simeq B$ one finds that $l_s$ is
a few tens of nm.

While appealing, theoretical approaches building on Eq.~(\ref{theory})
commonly rely on several approximations in order to make progress.
Analytical models usually schematise DNA as
a structureless polymer (or assume that the polymer configuration changes on a timescale much quicker than that of the protein movement~\cite{Metzler}), and also neglect {\it intersegmental transfers}, whereby 
the protein moves directly (i.e. without a 3D excursion) between two DNA 
regions which are close in 3D space, but can be far apart along the
DNA backbone. On the other hand, 
simulations~\cite{Langowski,Kafri1,Florescu,Gerland,Giulia}, 
usually treat the DNA as frozen (an exception is the lattice
study in~\cite{Gerland}), and disregard the base pair sequence of
DNA. 

Here we present a coarse grained simulation of the search process where 
we relax these two drastic approximations: we include the dynamics of all
components (DNA and proteins), and we consider a heterogeneous DNA.
We find that both aspects are crucial players in determining how fast 
facilitated diffusion is. First, we analyse the search process on a 
{\it string of rosettes}, which better represents the conformation of 
prokaryotic DNA as inferred from experiments~\cite{bacterialchromosome,peter}. 
We find that the relative position of the target with respect to the 
network may change $\tau_s$ by orders of magnitude. This giant effect
cannot be captured by the theory in Eq.~(\ref{theory}), in which
the target placement is immaterial. These findings
suggest that by changing the local DNA conformation it should be possible
to silence or express a given gene.
Second, if the DNA--protein
interaction is sequence-dependent~\cite{Mirny}, in general 
this slows down facilitated diffusion. However, through
a careful design of the DNA sequence, we show that
one can create a diffusional ``funnel'' that drives 
the protein to its target much more quickly.


In this work we used Brownian dynamics (BD) simulations in which we coarse grained DNA as a bead-and-spring polymer. Each of the $N$ beads in the DNA had a diameter $\sigma\sim 2.5$ nm, and neighbouring beads were connected by finitely extensible nonlinear elastic (FENE) springs. Proteins were modelled as spherical particles with a diameter of 3$\sigma$, with a spherical patch 
of radius $\sigma$ centred $1.1\sigma$ away from the protein centre (Fig. 1A). Only the latter was sticky for the DNA, via a (truncated) Lennard-Jones (LJ) interaction. All other coarse-grained beads interact via a purely repulsive potential which captures steric effects (this is achieved by truncating a LJ potential at a mutual distance of $2^{1/6}\sigma$). 
Finally, three neighbouring beads along the DNA are subjected to an additional force which models DNA semi-flexibility. Such a force comes from the gradient of the Kratky-Porod potential~\cite{kp}; this can be expressed as $K \cos{\theta}$, where $K=k_BTl_p/\sigma$ ($l_p=20$ $\sigma$ for DNA), and $\theta$ is the angle between the three neighbouring beads. 

We will refer to the full potential, including both LJ, Kratky-Porod and FENE terms, as $U$. If we denote the position of the $i$-th sphere in the simulation as ${\mathbf {x}_i}$, its evolution is determined by the following Langevin equation, 
\begin{equation}
m_i\frac{d^2 {\mathbf x}_i}{dt^2}=-\gamma_i\frac{d{\mathbf x}_i}{dt}-\nabla_i U+
\sqrt{2k_BT\gamma_i} {\boldsymbol \xi}_i(t), 
\end{equation}
where $\gamma_i$ is the friction felt by the particle, $\nabla_i=\frac{\partial}{\partial {\mathbf {x}_i}}$, $k_B$ is the Boltzmann constant, $T$ is the temperature,
$m_i$ is the mass of the $i-$th bead, and ${\boldsymbol \xi}_i(t)$ is an uncorrelated Gaussian noise with zero mean and unit variance~\cite{parameters}. All simulations were performed via the LAMMPS code~\cite{lammps}.


\begin{figure}
\begin{center}
\includegraphics{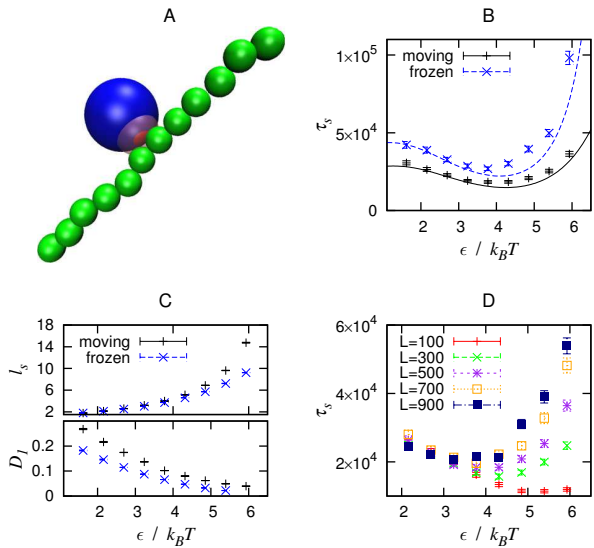}
\end{center}
\caption{
(A) Snapshot of a DNA segment and a model protein. (B) Mean search time $\tau_s$ for frozen and 
mobile DNA, as a function of DNA--protein affinity, $\epsilon$. Parameters are 
$L = 500\sigma$, $V \simeq 50000\sigma^3$ (the fraction of the volume occupied 
by the DNA is therefore $\simeq 1\%$), while data were averaged over 500 search
runs. The lines are a fit of the data with Eq.~(\ref{theory})~\cite{notefits}. 
(C) Plots of the 1D diffusion coefficient, $D_1$, and the sliding length, as a 
function of $\epsilon$. (D) Dependence of $\tau_s$ on affinity for various DNA lengths,
 $L$ (in units of $\sigma$), with fixed $V \simeq 50000\sigma^3$.}
\end{figure}

Fig. 1B shows the mean search time as a function of the DNA--protein affinity
$\epsilon$ (the depth of the attractive LJ potential, measured in units of 
$k_BT$), for the cases in which the DNA is either frozen (into a randomly 
chosen equilibrium configuration) or mobile. 
Since the sliding length increases with $\epsilon$ (Fig. 1C), our results are consistent with Eq.~(\ref{theory}) but now there is an optimal value $\epsilon^*$ which minimises $\tau_s$. Unlike in the theory we also observe a dependence of $D_1$ on $\epsilon$ (Fig. 1C), which comes from the presence of energy barriers felt by the protein while sliding -- in our case these are mainly due to the granularity of our polymer description, but they are likely
to be present for real DNA as well, due to the modulations in the major and 
minor grooves, and the curvature of the DNA. Experimentally $D_1$ has been shown to vary over a large range of values for different conditions, and DNA sequences~\cite{Wang}. 

Intriguingly, freezing the DNA leads to a much slower search, especially for large $\epsilon$. Our simulations also show that increasing the amount of genome available in the search volume, i.e. increasing $L$ at a fixed $V$, hinders, 
rather than helps, facilitated diffusion, unless the affinity is very small
(Fig. 1D). While the latter effect can be readily predicted from 
Eq.~(\ref{theory}), understanding the difference between the frozen and moving DNA cases requires a more detailed analysis of the protein trajectories in our numerical experiment. As one might expect, the 3D search time, $\tau_{\mathrm{3D}}$, which is dominant for small $\epsilon$, is larger (a $\sim40\%$ difference) for the frozen DNA; however we also observe an almost 2-fold larger value of $\tau_{\mathrm{1D}}$ for the frozen case.  Fig. 1C shows that
while $l_s$ is similar for the cases of mobile and frozen DNA, $D_{1}$
changes significantly, i.e. it is smaller in the case of frozen DNA. 
We ascribe this difference to the fact that, when mobile, the DNA  is able to
adjust locally to the presence of the protein, and hence can smooth out
some of the energy barriers which slow down the 1D sliding.
Once the measured values of $l_s$, $D_{1}$, $A$ and $B$~\cite{notefits} are
put into Eq.~(\ref{theory}), this actually provides a good fit to our data,
for both mobile and frozen DNA, as shown in Fig. 1B. The small residual
error may arise from the presence of the previously
mentioned ``intersegmental transfers'', which are neglected by the theory
-- indeed their presence somewhat changes the meaning of $l_s$. While traditionally $l_s$ is the length over which the protein ``slides'' during each encounter with the DNA, we here define it simply as the number of distinct DNA bead visited during the encounter --- whether consecutive along the contour, or separated due to intersegmental transfers.
Such events are present in our
simulations, and are more common in the mobile DNA case. 

\begin{figure}
\begin{center}
\includegraphics{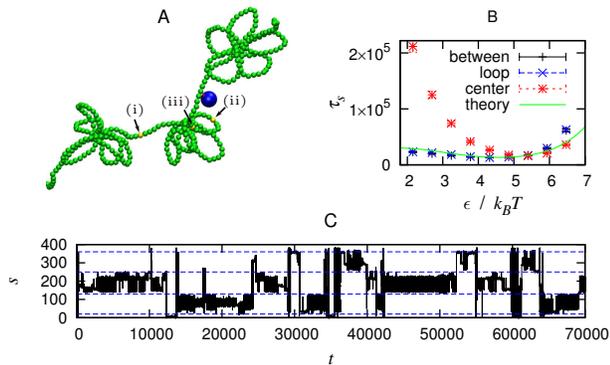}
\end{center}
\caption{(A) Snapshot of a string of rosettes. (B) Mean search time for a protein on a string of 3 rosettes with 5 loops (each of length $20\sigma$) on a DNA of length $L=380\sigma$ with $V\simeq 36000\sigma^3$ (1\% DNA volume fraction), for different
positions of the target: (i) between rosettes (bead 130), (ii) in the middle of a loop in a rosette (bead 190), and 
(iii) in the centre of a rosette (bead 180), as indicated in (A). The curve is the fit to the theory
in Eq.~\ref{theory}~\cite{notefits}. (Choosing a different number and size
of loops leads to qualitatively similar results.)
(C) Time series of the DNA-bead ($s$) nearest to the protein
at a given instant, showing trapping close to a rosette centre ($\epsilon=5.9k_BT$). Dashed lines separate beads belonging to different rosettes.}
\end{figure}

The DNA conformations found {\it in vivo} in bacteria, 
while not yet well characterised, are likely 
to be quite far from those of the self-avoiding polymer 
normally considered in the theories for this process, and which we studied
in Fig. 1. Within the prokaryotic cytosol, DNA is known to be highly looped,
due to the presence of DNA-binding architectural proteins such as condensins --
this helps to achieve the compaction which is required to fit
the whole genome within the narrow volume of a single cell.
Therefore we consider in Fig. 2 the dynamics of a protein searching for its target
on a DNA which is made up of a string of rosettes, each of which consists of
a series of loops joined together (see Fig. 2A).
This idealised conformation gives a realistic
local view of bacterial DNA according to a number of biological models (see e.g.~\cite{peter}) 
and is simple enough to be included in our modelling. 

Fig. 2B shows the mean search time $\tau_s$ as a function of $\epsilon$, for
three different target positions: (i) in the centre of a rosette, (ii) in the middle of a loop
in a rosette, and (iii)  between rosettes. Our results show that when the  affinity 
between the protein and the DNA is small, so that 3D diffusion dominates over 1D diffusion during the search, it takes much longer to find a target in the centre of a rosette. 
Such a target is
more difficult to reach as the surrounding loops are in the way. Interestingly,
this trend reverses for larger values of the affinity. To understand this, we observe that in the large $\epsilon$ regime 
each of the rosettes acts as a trap for the protein, 
i.e. it spends a large amount of time in a rosette, before moving to another one (see Fig. 2C). 
Since sliding is the dominant transport mechanism, rather than acting as a shield, the loops allow the protein to slide into the centre of the rosette. Once there intersegmental transfers are more likely to
keep the protein near that centre than take it elsewhere. Such a mechanism then renders it
easier to find the target if it is close to one of the traps.

Fig. 2 therefore demonstrates that DNA topology and target positioning, together with DNA--protein
affinity, can be used to control the relative ease with which different regions of the genome can
be accessed by proteins. We highlight that this conclusion is outside the scope of most facilitated 
diffusion theories based on arguments such as that in Eq.~(\ref{theory}), in which the position of the target does not feature.
More quantitatively, we have computed $\tau_{\mathrm{3D}}$ and $\tau_{\mathrm{1D}}$,
as well as $D_{1}$, and $l_s$ from our data, and found that while $\tau_{\mathrm{3D}}\sim V/D_3 L$
still holds, it is not possible to fit $\tau_{\mathrm{1D}}$ to the functional form $Bl_s^2/D_1$ throughout
the $\epsilon$ range considered here (not shown). This is because
the rosette structure introduces large correlations between the points where the protein leaves and rejoins the DNA for each 3D excursion, meaning that
$N_s$ is very sensitive to the target position and poorly predicted by 
Eq.~(\ref{theory}) (see Fig. 2B).

\begin{figure}
\begin{center}
\includegraphics{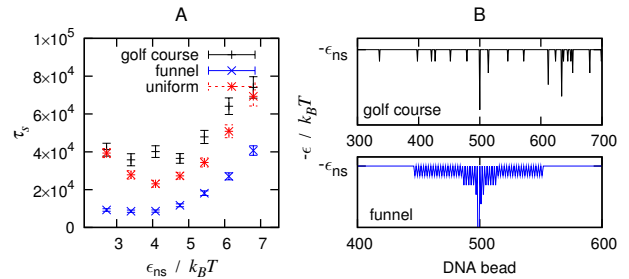}
\end{center}
\caption{(A) Plot of $\tau_s$ as a function of $\epsilon_{\rm ns}$ for three DNA sequences:
(i) homogeneous DNA, (ii) randomly positioned traps and (iii) traps clustered so
as to provide a funnel towards the target. (B) Schematic showing the DNA--protein affinity for each bead in a typical section of DNA, for the two different trap arrangements. In the funnel case traps alternate with non-specific beads. In each case the target is at bead 500, $L=1000$ and $V\simeq10^5\sigma^3$ (1\% DNA volume fraction).
}
\end{figure}

We now turn to the discussion of another aspect found in real DNA and commonly neglected in theoretical
work: sequence heterogeneity. The DNA sequence leads to a non-uniform free energy landscape for a protein
sliding along it. In order to describe such a landscape, we allow the DNA--protein interaction to vary from
one DNA bead to another, with the bead-dependent affinity set as prescribed by the model proposed in
Ref.~\cite{GerlandPNAS}. There it was postulated that there exists 
two possible states for a protein attached to the genome: it
can either bind in a non-specific mode -- with constant affinity $\epsilon_{\rm ns}$, 
or in a sequence-dependent, specific, mode -- with affinity larger than $\epsilon_{\rm ns}$. The model in \cite{GerlandPNAS} assumes that the two states are 
in equilibrium, so the protein will be found in whichever state offers a stronger interaction. 
In our simulations, for DNA bead $s$ we choose a specific interaction strength $\epsilon_{\rm s}(s)$ according to an appropriate distribution~\cite{Voituriez,GerlandPNAS}; the affinity for that bead is then taken to be whichever is the larger of $\epsilon_{\rm s}(s)$ and $\epsilon_{\rm ns}$. In 
practice this leads
to a free energy profile with most beads favouring the nonspecific interaction 
strength $\epsilon_{\rm ns}$, with a small number of ``traps'' with a greater interaction energy.
Unlike those of the rosettes considered in Fig. 2, which are determined by the 3D structure
of the DNA, such traps are encoded in the 1D sequence of bases.

Fig. 3 shows the dependence of $\tau_s$ on $\epsilon_{\rm ns}$
for a DNA chain with $L=1000\sigma$ 
(corresponding to $\sim 7350$ base pairs), where 
the trap strength and number of traps have been determined on the basis of the statistics for the binding of
a ``typical'' bacterial transcription factor (TF)~\cite{Voituriez,Mirny,notesequencechoice}. 
We focused on the case in which the target:protein
interaction energy is larger than the affinity with any of the traps,
which is the most common for real TFs~\cite{Voituriez}. 
We compared the case of homogeneous DNA with a nonspecific interaction $\epsilon_{\rm ns}$, with two inhomogeneous sequences: 
(i) that in which the position of the traps is random, leading to a ``golf-course'' free energy landscape; and 
(ii) that in which the DNA sites with enhanced affinity for the proteins are clustered around the target (alternating non-specific and enhanced binding beads)
so as to provide a potential funnel driving the protein to it (see Fig. 3B). 
We refer to these two situations 
as the golf-course and funnel case respectively.

A general Kramers' argument suggests that the time the protein spends in a trap may be estimated as 
$\tau_{\rm trap}=\tau_0e^{(\epsilon_{\rm trap}-\epsilon_{\rm ns})/k_BT}$, where $\tau_0\sim \sigma^2/D_1$ is the time it 
takes a protein to move from one non-specifically interacting DNA bead to the next. It is therefore not surprising
that this case leads to a far larger mean search time with respect to the homogeneous DNA case,
where the binding of the protein to the genome is always ``non-specific'' (see Fig. 3A). If the search 
involved 1D sliding along the DNA contour alone, one might expect that if the non-specific interaction 
$\epsilon_{\rm ns}$ were increased at fixed $\epsilon_{\rm trap}$ 
then this would lead to an exponential decrease in the search time (in line with
the decrease in trap depth); however, for facilitated diffusion, this is balanced by the increase in 
$l_s$ (above its optimum value) which leads to a slower search.

The Kramers' argument does not apply to the funnel case, which eliminates
traps other than near the target -- intersegmental transfers from one ``trap''
to the next provide an alternative transport mechanism which avoids 
slowdown due to the rugged 1D potential.
One may then expect that $\tau_s$ should be similar to the one observed with
uniform DNA, with some enhancement due to the binding gradient which drives 
the protein towards the target once it is in its close proximity. 
Strikingly, the speed up with respect to the uniform case may instead reach about one order of magnitude
(and more than two with respect to the golf-course case). This is probably because the presence of the funnel
can decrease the likelihood of the protein being transported away from the vicinity of the target, 
even for small $\epsilon_{\rm ns}$~\cite{noteshuffledfunnel}.

The dramatic difference between search efficiency in the golf-course and funnel case is 
a consequence of the assumption (from~\cite{GerlandPNAS}) that proteins can bind to
DNA either non-specifically or specifically, and the two states are in thermodynamic equilibrium so that
the optimal binding for each site can be selected quickly. It is currently not clear whether this is a 
correct assumption -- an alternative 
suggestion~\cite{Mirny,Kafri2}
is that what matters may be the energy barrier between the specific and nonspecific bound states, rather than their
absolute binding energy. If the energy barrier between the states was very large for all sites except the target, 
then our funnel sequence should not lead to much enhancement in the efficiency with respect to the random case. 
That is to say, the protein would see only a flat (non-specific) landscape irrespective of the sequence, 
and the ``funnel'' would not be accessible to it. It would therefore be interesting to perform
{\it in vitro} single molecule experiments analogous to those of Ref.~\cite{Xie}, where the DNA sequence is 
either random or designed so as to create the funnel we considered in Fig. 3.  In this way one may directly test whether
the predictions from our simulations hold, and hence determine which of the two theories mentioned above for 
DNA--protein binding applies in reality.

In conclusion, we have presented Brownian dynamics simulations of 
the facilitated diffusion of a protein on DNA. Unlike previous numerical work, 
we have focused on the impact of 3D DNA conformation  
and sequence heterogeneity on the search dynamics.
We have found that the presence of loops in the DNA may provide a way to tune the accessibility of a 
target on the genome, which cannot be accounted for by existing analytical theories. 
By considering a string of rosettes for the DNA conformations, we have seen that when the 
target is in the centre of a rosette and the DNA--protein affinity is small, the time needed to 
find it is larger than in the case when the target is positioned between rosettes. 
This effect reverses for high affinity -- in this
regime each of the rosettes acts as a configurational trap, in the vicinity of which the protein
lingers for a long time.  While the conformation of prokaryotic genomes may 
adopt far more complicated topologies than
the string of rosettes which we have considered, our results are generic 
in predicting a dependence on the
relative positioning of loops and targets. Hence we expect they should also apply to more
disordered loop networks.
Finally, we have considered the case of a heterogeneous DNA, where the affinity between genome and protein
is site-dependent, thereby introducing traps in the facilitated diffusion of the protein.
When the sequence is random, these traps severely slow down the search process. 
However, when the sequence is designed so as to provide a funnel-like landscape around the target, the
search may become much faster. Experiments to test this latter prediction should lead to a better understanding of the way proteins bind to DNA.

We acknowledge EPSRC grant EP/I034661/1 for funding. MEC is funded by the Royal Society.


\begin{thebibliography}{99}
\bibitem{notesearchrounds} This relies on the assumption that the
point on the DNA where the protein starts a slide is not correlated with the last point
of its previous interaction with the genome.
\bibitem{parameters} In our simulations we fixed the following parameters: $m_i=1$ for a DNA-bead, and $m_i=3$ for a protein (core plus patch); likewise $\gamma_i=1$ for a DNA-bead and $\gamma_i=3$ for a protein. The latter choice defines the viscosity $\eta$ of the fictitious Brownian fluid, and the diameter $d$ of the particles, via the usual Stokes-Einstein relation $\gamma=3\pi\eta d$. 
We use a cubic simulation box with periodic boundary conditions (this avoids complications due to interactions between the DNA or proteins and the boundaries). Systems are equilibrated for at least $10^6$ time steps before the attractive interactions are switched on, so the initial DNA configuration and protein position are random. To map timescales to physical units, 
we note a time unit in Figs. 1-3 corresponds to 36 ns
(if $\eta=1$ cP).
\bibitem{notefits} To fit our data with the theory in Eq.~(\ref{theory}), we
need to estimate the constants $A$ and $B$, by finding $\tau_{\mathrm{3D}}$ and $\tau_{\mathrm{1D}}$ from the numerical trajectories. From the known values of $V$, $L$ and $D_3$ we fit the $\tau_{\mathrm{3D}}$ data to find $A$. From a Kramers' argument we expect that $\tau_{\mathrm{1D}}$ will depend exponentially on the DNA--protein affinity $\epsilon$; fitting gives a function $\tau_{\mathrm{1D}}(\epsilon)$. To find $N_s$ and the constant $B$ we need to know $D_1$ and $l_s$, both of which are found to depend on $\epsilon$. To estimate the 1D diffusion we use the statistics of the sliding events; i.e. we calculate the diffusion constant for each value of $\epsilon$ from the mean square 1D displacement (along the DNA) versus time, and then fit to find a function $D_1(\epsilon)$. Finally, $B$ is estimated by using the previously found expressions for $\tau_{\mathrm{1D}}(\epsilon)$ and $D_1(\epsilon)$, and fitting $l_s$ (measured from the trajectories) to the equation $l_s=B^{-1/2}\sqrt{\tau_{\mathrm{1D}}(\epsilon)D_1(\epsilon)}$.
\bibitem{noteshuffledfunnel}
This interpretation is
corroborated by further simulations in which we shuffled the position of
the traps within the funnel region. Fewer search rounds are required to find the target than in the uniform case, and the search time is always less than the golf course case. However, the gain is less than the unshuffled
funnel as this configuration provides no binding energy gradient.
\end{thebibliography}
\end{document}